\documentclass[sigconf, nonacm]{acmart}
\usepackage{tikz}
\usetikzlibrary{calc}
\usetikzlibrary{patterns}
\usetikzlibrary{decorations.pathreplacing}
\usepackage[linesnumbered,ruled,vlined]{algorithm2e}
\usepackage{adjustbox}
\usepackage{booktabs}
\theoremstyle{acmdefinition}
\newtheorem*{remark}{Remark}
\AtBeginDocument{%
  \providecommand\BibTeX{{%
    \normalfont B\kern-0.5em{\scshape i\kern-0.25em b}\kern-0.8em\TeX}}}

\copyrightyear{2021}
\acmYear{2021}
\setcopyright{none}
\acmConference[ISPD '21]{Proceedings of the 2021
International Symposium on Physical Design}{March 22--24, 2021}{Virtual
Event, USA}
\acmBooktitle{Proceedings of the 2021 International Symposium on Physical
Design (ISPD '21), March 22--24, 2021, Virtual Event, USA}
\acmPrice{15.00}
\acmDOI{10.1145/3439706.3447044}
\acmISBN{978-1-4503-8300-4/21/03}

\begin{document}
\fancyhead{}
\title{A Fast Optimal Double Row Legalization Algorithm}


\author{Stefan Hougardy}
\email{hougardy@or.uni-bonn.de}
\affiliation{%
  \institution{Research Institute for Discrete Mathematics, University of Bonn}
  \streetaddress{Lenn\'estr. 2}
  \city{Bonn}
  \state{Germany}
  \postcode{53113}
}

\author{Meike Neuwohner}
\orcid{0000-0002-3664-3687}
\email{neuwohner@or.uni-bonn.de}
\affiliation{%
  \institution{Research Institute for Discrete Mathematics, University of Bonn}
  \streetaddress{Lenn\'estr. 2}
  \city{Bonn}
  \state{Germany}
  \postcode{53113}
}

\author{Ulrike Schorr}
\email{uschorr@cadence.com}
\affiliation{%
  \institution{Cadence Design Systems Inc.}
  \streetaddress{Mozartstr. 2}
  \city{Munich}
  \state{Germany}
  \postcode{85622}
}


\begin{abstract}
 In Placement Legalization, it is often assumed that (almost) all standard cells possess the same height and can therefore be aligned in \emph{cell rows}, which can then be treated independently. However, this is no longer true for recent technologies, where a substantial number of cells of double- or even arbitrary multiple-row height is to be expected. Due to interdependencies between the cell placements within several rows, the legalization task becomes considerably harder. In this paper, we show how to optimize quadratic cell movement for pairs of adjacent rows comprising cells of single- as well as double-row height with a fixed left-to-right ordering in time $\mathcal{O}(n\cdot\log(n))$, whereby $n$ denotes the number of cells involved. Opposed to prior works, we thereby do not artificially bound the maximum cell movement and can guarantee to find an optimum solution. Experimental results show an average percental decrease of over $26\%$ in the total quadratic movement when compared to a legalization approach that fixes cells of more than single-row height after Global Placement.
\end{abstract}




\maketitle

\section{Introduction}
The \emph{Standard Placement Problem} captures the task of locating hundreds of thousands or even millions of standard cells, which are usually assumed to exhibit uniform heights, within the rectangular chip area. Thereby, multiple objectives such as minimizing the total length of inter-cell electrical connections (nets) or achieving desirable timing properties have to be respected. Given the fact that even the underlying packing problem is strongly $NP$-hard \cite{GareyJohnson}, the placement task is most commonly split into the three sub-problems of \emph{Global Placement}, \emph{Legalization} and \emph{Detailed Placement}. Global Placement aims at finding cell locations that approximately minimize the total netlength for a certain net model and obey bounds on local packing density, but does not have to ensure internal disjointness of shapes. The Legalization step deals with resolving the remaining overlaps by shifting cells locally, trying to minimize either netlength or the total (squared) cell displacement. The latter is desirable because it honors the quality of the Global Placement result (e.g. w.r.t. timing) and balances cell movement. Detailed Placement usually incorporates several post-optimization routines.\\
When only cells of single-row height are present, the Standard Cell Legalizers ``Tetris'' \cite{hill} and ``Abacus'' \cite{Abacus} produce good results. They process the cells one by one, ordered by the $x$-coordinates of their Global Placement positions, and place each cell at the closest free position \cite{hill} or at the end of a nearby row, choosing the one that allows for the minimum possible total cell movement \cite{Abacus}. Another strategy, which is employed within the \mbox{BonnTools} project \cite{BonnTools},\cite{LegalizationBonnTools}, uses a min-cost-flow approach to first assign the cells to \emph{zones}, unblocked parts of a row \cite{BrennerMCF}. Fixing the left-to-right ordering of the cells contained within each zone to the one imposed by the Global Placement locations, legal cell positions are then obtained by minimizing the total squared cell displacement (or (weighted) bounding box netlength) within each zone. The latter task is captured by the \emph{Single Row Problem}, which also occurs as a sub-problem of the Abacus Legalizer. It was first studied by Kahng, Tucker and Zelikovsky \cite{KahngTuckerZelikovsky}, who suggested the \emph{Clumping Algorithm} to tackle it. While their implementation runs in $\theta(m\cdot\log^2(m))$ for unit net weights (where $m$ denotes the number of nets), the fastest implementation, which is due to Suhl \cite{Suhl}, achieves a running time of $\mathcal{O}(m\cdot\log(m))$ even for general net weights. A similar result has been obtained in the context of scheduling \cite{SchedulingGareyTarjanWilfong}. When the goal is to optimize quadratic cell movement, the Clumping Algorithm can easily be implemented to run in time linear in the number of cells.\\
While the mentioned approaches work well in the presence of uniform cell heights, it is not obvious how to generalize them to a setting where cells of double- or even arbitrary multiple-row height may occur. Wang et al.\ \cite{EffectiveLegalizationAlgorithm} try to adapt the Clumping Algorithm to the double-row case, but manage to guarantee optimality only in a very restricted setting. In contrast to this, Wu and Chu \cite{WuChu} suggest to handle cells of double-row height by, depending on the placement density, either inflating or matching cells of single-row height to ensure uniform cell heights again. However, as was already pointed out in \cite{MrDP}, this strategy can neither handle distinct power alignment constraints nor cells covering more than two rows. Besides, both merging and inflating cells may drastically reduce the placement flexibility as well as lead to a significant area overhead.\\Many other authors, therefore, settle for a dynamic programming solution instead of generalizing the Clumping Algorithm, guaranteeing a reasonable runtime by artificially bounding the maximum displacement allowed for each cell by a small number of placement sites. In exchange, they show how to make their dynamic program aware of several other desirable objective traits or incorporate a larger degree of freedom by allowing for a local reordering of cells, even between multiple rows \cite{SpacingRules}, \cite{DiffusionSteps}, \cite{DiffusionStepsEnhanced}, \cite{MrDP}.\\
Other approaches comprise solving a linear complementarity problem to approximately minimize the squared cell movement and then resolving the remaining overlaps \cite{QPLegalization2017}, \cite{Analytical}, \cite{QPLegalization2018}, applying integer linear programming to legalize sufficiently small regions of the chip separately \cite{hung2017mixed}, or making use of a cell insertion scheme \cite{ChowPuiYoung}, combined with bipartite matching and min-cost-flow-algorithms~\cite{Routability}.\\
 In this paper, we present a fast $\mathcal{O}(n\log n)$-time (where $n$ denotes the number of cells) algorithm minimizing the total quadratic displacement for cells of single- and double-row height that need to be accommodated in two adjacent rows obeying a fixed ordering of the cells covering each row. In contrast to previous dynamic programming approaches, we do not need to artificially restrict the number of available positions for each cell, which may be beneficial for regions of low density and when dealing with coarser grid sizes for double-row cells, which our algorithm can take into account. Moreover, our approach can be extended to support rectangular movebounds for the cells.\\
 The rest of this paper is organized as follows: In Section~\ref{SecPrelim}, we discuss the Single Row Problem, the Clumping Algorithm and its implementation for \emph{piecewise quadratic} cost functions. In Section~\ref{SecDoubleRowProblem}, we then introduce the \emph{Double Row Problem} and show how to reduce it to the Single Row Problem in Section~\ref{SecReduction}. Finally, Section~\ref{SecPracticalResults} presents our experimental results.
 \section{Preliminaries}\label{SecPrelim}
 The following section comprises the base results our reduction from the Double Row to the Single Row Problem builds upon.
 \begin{itemize}
  \item Section~\ref{SecSingleRowAndClumping} reviews the Clumping Algorithm and its analysis.
  \item Theorem~\ref{TheoKeepBlocks} points out how an optimum solution to the Single Row Problem changes when the domain is restricted.
  \item Section~\ref{SecImplClumping} discusses an efficient implementation of the Clumping Algorithm for piecewise quadratic cost functions.
 \end{itemize}

 \subsection{The Single Row Problem and the Clumping Algorithm\label{SecSingleRowAndClumping}}
 \begin{definition}[Single Row Problem]
 	\begin{description}
 		\item[]
 		\item[Instance: ]
 		A tuple $(\mathcal{C},w,x_{min},x_{max},(f_i)_{i=1}^n)$ consisting of
 		\begin{itemize}
 			\item a set $\mathcal{C}:=\{C_1,\dots,C_n\}$ of cells, 
 			\item cell widths $w:\mathcal{C}\rightarrow\mathbb{R}^+$,
 			\item a minimum and  maximum coordinate $x_{min},x_{max}\in\mathbb{R}$\\ satisfying $\sum_{i=1}^n w(C_i)\leq x_{max}-x_{min}$ and
 			\item convex, continuous functions $f_i:\mathbb{R}\rightarrow\mathbb{R}$ for $i=1,\dots,n$.
 		\end{itemize}
 		\item[Task: ] Find coordinates $(x_i)_{i=1}^n$ minimizing $\sum_{i=1}^n f_i(x_i)$ subject to
 		\begin{itemize}
 			\item $x_{min}\leq x_1,$
 			\item $x_i+w(C_i)\leq x_{i+1}$ for $i=1,\dots,n-1$ and 
 			\item $x_n+w(C_n)\leq x_{max}$.
 		\end{itemize}
 	\end{description}
 	For $i=1,\dots,n$, we write \[[f_i^-,f_i^+]:=\mathrm{argmin}\{f_i(x),x\in[x_{min}+\sum_{j=1}^{i-1}w(C_j),x_{max}-\sum_{j=i}^n w(C_j)]\}.\]
 \end{definition}
The Single Row Problem can be solved by the aforementioned \emph{Clumping Algorithm} \cite{KahngTuckerZelikovsky}. The 
given formulation of the Clumping Algorithm (Algorithm~\ref{ClumpingAlgo}) is based on \cite{BrennerVygen}.

\begin{algorithm}
\DontPrintSemicolon
	\KwIn{An instance of the Single Row Problem given by\;
		an ordered list $\mathcal{L}=(C_1,\dots,C_n)$ of cells,\;
	cell widths $w:\{C_1,\dots,C_n\}\rightarrow\mathbb{R}^+$,\;
	a row interval $[x_{min},x_{max}]$ and\;
convex cost functions $(f_i)_{i=1}^n$.}
\KwOut{Optimum positions $(x_i)_{i=1}^n$.}
Add an auxiliary element $C_0$ to the front of $\mathcal{L}$ and set $x_0\gets x_{min}$ and $w_0\gets 0$.\;
\For{$i\gets 1$ \KwTo $n$}{
	Compute $f_i^-$ and $f_i^+$.\;
	$w_i\gets w(C_i)$\;
}
\For{$i\gets 1$ \KwTo $n$}{
$PLACE(C_i,\mathcal{L})$\;}
\For{$i\gets 1$ \KwTo $n$ $\mathrm{with}$ $C_i\not\in\mathcal{L}$}{
$x_i\gets x_{i-1}+w(C_{i-1})$\;}
\Return $(x_i)_{i=1}^n$\;
\caption{Clumping Algorithm}\label{ClumpingAlgo}	
\end{algorithm}
\begin{algorithm}
\DontPrintSemicolon
	$C_h\gets$ predecessor of $C_i$ in $\mathcal{L}$\;
	\If{$x_h+w_h\leq f^+_i$}{
	$x_i\gets \max\{x_h+w_h, f^-_i\}$\;}
\Else{
$COLLAPSE(C_h,C_i,\mathcal{L})$\;
$PLACE(C_h,\mathcal{L})$\;}
\caption{$PLACE(C_i,\mathcal{L})$}	
\end{algorithm}
\begin{algorithm}
\DontPrintSemicolon
	Redefine $f_h$ as $x\mapsto f_h(x)+f_i(x+w_h)$ and update $f_h^-$ and $f_h^+$ (w.r.t. $[x_{min}+\sum_{j=1}^{h-1}w(C_j), x_{max}-\sum_{j=h}^n w(C_j)]$)\;
	$w_h\gets w_h+w_i$\;
	Remove $C_i$ from $\mathcal{L}$\;
	\caption{$COLLAPSE(C_h,C_i,\mathcal{L})$}
\end{algorithm}
\begin{theorem}[\cite{KahngTuckerZelikovsky}]
	The Clumping Algorithm finds an optimum placement. \label{TheoClumpingAlgoOpt}
\end{theorem}
We prove a slightly stronger statement which we will need at a later point. In order to formulate it, we have to introduce the notion of a \emph{block}, which we define as follows:
For a cell $C_i\in\mathcal{L}$, the block $B(i)$ represented by $C_i$ is defined to be the consecutive set of cells $B(i):=\{C_j:i\leq j\leq n\wedge\not\exists C_l\in\mathcal{L}:i<l\leq j\}$. The blocks present at a given point during the run of the Clumping Algorithm indicate sets of cells that the algorithm forces to be placed contiguously (or has \emph{clumped together}) at that time. Note that the partition into blocks can only get coarser throughout the run of the algorithm. 
\begin{theorem}
 Let $I':=(\mathcal{C},w,x'_{min},x'_{max},(f_i)_{i=1}^n)$ be an instance of the Single Row Problem, let $x_{min}\leq x'_{min}<x'_{max}\leq x_{max}$ and let $I$ denote the instance of the Single Row Problem that arises from replacing $x'_{min}$ and $x'_{max}$ by $x_{min}$ and $x_{max}$, respectively. Then there exists an optimum solution $(x^*_i)_{i=1}^n$ for $I'$ such that for any block $B(i)$ formed during the run of Algorithm~\ref{ClumpingAlgo} on $I$, the cells in $B(i)$ are placed contiguously.\label{TheoKeepBlocks}  
\end{theorem}
\begin{proof}
 By induction on the number of calls to $COLLAPSE$. Initially, the statement is clearly true because every cell constitutes a block on its own. Consider a call to $COLLAPSE$ where two blocks $B(h)$ and $B(i)$ are united by deleting $C_i$ from $\mathcal{L}$, and pick an optimum solution $(x^*_i)_{i=1}^n$ for $I'$ respecting all previously formed blocks. If additionally $x^*_{i-1}+w(C_{i-1})=x^*_i$, we are done, so assume $x^*_h+\sum_{l=h}^{i-1} w(C_l)= x^*_{i-1}+w(C_{i-1})<x^*_i$. By construction of the algorithm, we have $f_h^-\leq x_h\leq f_h^+$, $w_h=\sum_{l=h}^{i-1} w(C_l)$ and $x_h+w_h>f_i^+$. If $x^*_i>f^+_i$, then we can shift $B(i)$ to the left until it hits $\max\{x^*_{i-1}+w(C_{i-1}),f_i^+\}$ and thereby decrease the total cost since the cost function $f_i$ of $B(i)$ is strictly  monotonically increasing on $[f_i^+,x_{max}-\sum_{j=i}^n w(C_j)]\supseteq[f_i^+,x'_{max}-\sum_{j=i}^n w(C_j)]$, a contradiction to the assumed optimality of $(x^*_i)_{i=1}^n$. Hence $x^*_i\leq f_i^+$. Then $x^*_i-w_h<x_h\leq f_h^+$, so we can shift $B(h)$ to the right until it hits the left boundary of $B(i)$ without increasing the total cost since the cost function $f_h$ of $B(h)$ is monotonically decreasing on $[x_{min}+\sum_{j=1}^{h-1}w(C_j),f_h^+]\supseteq[x'_{min}+\sum_{j=1}^{h-1}w(C_j),f_h^+]$.
\end{proof}
\begin{remark}
 Together with the fact that the Clumping Algorithm places each block $B(i)$ with its optimum range $[f_i^-,f_i^+]$ and hence also within $[x_{min},x_{max}-w_i]$ (whereby $f_i$ and $w_i$ refer to the respective values after $B(i)$ has been formed), Theorem~\ref{TheoKeepBlocks} implies optimality and therefore in particular the correctness of Theorem~\ref{TheoClumpingAlgoOpt}.
\end{remark}
\begin{theorem}
 Let $I$ and $I'$ be as in Theorem~\ref{TheoKeepBlocks} and let $(x^*_i)_{i=1}^n$ be the solution computed by a run of the Clumping Algorithm on $I$. Then an optimum solution $(x'^{*}_i)_{i=1}^n$ for $I'$ is given by
 \[x'^{*}_i=\min\left\{x'_{max}-\sum_{j=i}^n w(C_j), \max\left\{x'_{min}+\sum_{j=1}^{i-1} w(C_j), x^*_i\right\}\right\}\] for $i=1,\dots,n$.\label{TheoOptSolutionShrinkInterval}
\end{theorem}
\begin{proof}
 Feasibility follows easily from the fact that we have $x'_{max}-x'_{min}\geq \sum_{i=1}^n w(C_i)$ by definition of the Single Row Problem. By Theorem~\ref{TheoKeepBlocks}, it, therefore, suffices to show that $(x'^{*}_i)_{i=1}^n$ places each block $B(i)$ arising from the run of the Clumping Algorithm on $I$ optimally. Pick such a block $B(i)$ and call its cumulated cost function to which $f_i$ is set during the course of the algorithm $\bar{f}_i$. Then by definition of the Clumping Algorithm, we have $x^*_i\in[\bar{f}_i^-,\bar{f}_i^+]$. We distinguish the three cases \begin{itemize}\item $x^*_i<x'_{min}+\sum_{j=1}^{i-1} w(C_j)$, \item $x_i^*\in[x'_{min}+\sum_{j=1}^{i-1} w(C_j),x'_{max}-\sum_{j=i}^n w(C_j)]$ and \item $x'_{max}-\sum_{j=i}^n w(C_j)<x_i^*$.\end{itemize} In the first case, $x'^*_i=x'_{min}+\sum_{j=1}^{i-1} w(C_j)$ is set to the leftmost feasible position and furthermore, $\bar{f}_i$ is monotonically increasing to the right of $x^*_i<x'^*_i$, showing that $B(i)$ is placed optimally. In the second case, $x'^*_i=x^*_i$ is placed within the optimum range of \mbox{$\bar{f}_i\upharpoonright[x_{min}+\sum_{j=1}^{i-1}w(C_j),x_{max}-\sum_{j=i}^{n}w(C_j)]$} and therefore in particular occupies an optimum position for this function. Finally, in the third case, we get $x'^*_i=x'_{max}-\sum_{j=i}^n w(C_j)$, which is the rightmost feasible position $C_i$ may attain. Given that $\bar{f}_i$ is monotonically decreasing on $[x'_{min}+\sum_{j=1}^{i-1}w(C_j),x'^*_i]\subseteq [x_{min}+\sum_{j=1}^{i-1}w(C_j), \bar{f}_i^+]$, optimality follows again.
\end{proof}
Note that if all of the $f_i$ are quadratic functions stored as triples $(a,b,c)$ of coefficients such that $f_i:x\mapsto a\cdot x^2+b\cdot x+c$, the Clumping Algorithm can be implemented to run in linear time, as pointed out, for example, in \cite{Suhl}, since the computation of minima as well as shifting a quadratic function in $x$-direction or adding it to another one only requires a constant number of arithmetic operations on the respective coefficients.

\subsection{Implementation of the Clumping Algorithm with piecewise quadratic objective functions\label{SecImplClumping}}
Our strategy to solve the problem of minimizing squared movement within two adjacent rows containing cells of both single- and double-row height with a prescribed left-to-right ordering is based on a reduction of an instance of the latter problem to an instance of the Single Row Problem with \emph{piecewise quadratic objective functions}.
In the following subsection, we therefore discuss how to implement the Clumping Algorithm in this case.
\begin{definition}[piecewise quadratic function]
 For $[a,b]\subseteq\mathbb{R}$, we call a continuous function $f:[a,b]\rightarrow\mathbb{R}$ \emph{piecewise quadratic} if there exist a nonnegative integer $k$ and \begin{itemize}\item real numbers $a=:x_0<x_1<\dots<x_k<x_{k+1}:=b$ and \item quadratic functions $(f_i:\mathbb{R}\rightarrow\mathbb{R})_{i=0}^k$\end{itemize} such that $f\upharpoonright[x_i,x_{i+1}]=f_i\upharpoonright[x_i,x_{i+1}]$ for all $i=0,\dots,k$. The positions $(x_i)_{i=1}^k$ are called \emph{kinks} of $f$. Note that there exists a unique representation of $f$ with $f_i\neq f_{i+1}$ for all $i=0,\dots,k-1$, to which we refer when talking about \emph{the} set of kinks of a piecewise quadratic function.
\end{definition}
Our goal is to achieve a running time of $\mathcal{O}((n+m)\log(\min\{n,m\}))$ for the Clumping Algorithm, where $n$ denotes the number of cells and $m$ specifies the total number of kinks occurring among all cost functions. Therefore, we suggest an implementation of the algorithm that is based on the one proposed in \cite{Suhl} for the case of piecewise linear objective functions. Due to page limit, we do not present a detailed description, but rather give a short overview of the data structures used as well as a brief outline of the analysis. 
\paragraph{Representation of cost functions of cells} We associate the quadratic function $x\mapsto a\cdot x^2+b\cdot 
x+c$ with the triple $(a,b,c)$ and store the restriction $f_i\upharpoonright[x_{min},x_{max}]$ of the piecewise 
quadratic cost function $f_i$ as follows:\\ Let $x_{min}=:p_i^{m_i+1}<p_i^{m_i}<\dots<p_i^1<p_i^0:=x_{max}$ such that 
$\{p_i^{1},\dots,p_i^{m_i}\}$ is the set of kinks of $f_i\upharpoonright[x_{min},x_{max}]$ and let 
\mbox{$f_i\upharpoonright[p_i^{j+1},p_i^j]$} be given by the quadratic function $f_i^j$, \mbox{$j=0,\dots,m_i$}. Then we 
represent $f_i$ by the ordered list \mbox{$F_i:=((p_i^{j+1},f_i^j))_{j=0}^{m_i}$} consisting of pairs of quadratic 
functions defining $f_i\upharpoonright[x_{min},x_{max}]$ on a certain interval and the left boundary of their domain. 
Throughout the algorithm, for each cell $C_i$ that has already been processed and is currently placed at the position 
$x_i$, we maintain the index $j(i)\in\{0,\dots,m_i\}$ for which $p_i^{j(i)+1}<x_i\leq p_i^{j(i)}$ respectively 
$j(i)=m_i$ if $x_i=x_{min}$. Observe that if we implicitly assume all cells to be located at $x_{max}$ initially 
and further consider a cell $C_j\in B(i)$ as being placed at $x_i+\sum_{l=i}^{j-1}w(C_l)$, cells never move to the 
right during a run of the Clumping Algorithm. To see this, note that by definition of $f_i^-$ and $f_i^+$, each cell 
is located within $[x_{min},x_{max}]$ by construction. Moreover, whenever $x_h$ is reassigned after a call to 
$COLLAPSE(C_h,C_i,\mathcal{L})$, then $h\neq 0$ and for $C_g$ the predecessor of $C_h$ in $\mathcal{L}$, we get 
$\max\{x_g+w_g,f_h^-\}= x_h\leq f_h^+$ and $x_h+w_h> f_i^+\geq f_i^-$ before $COLLAPSE$ is performed. Hence, after the update of 
$f_h$, we have $f_h^-\leq x_h$, implying that $x_h$ is decreased, remains unchanged or another call to $COLLAPSE$ is 
launched. In the first case, all already processed cells $C_i$ with $i> h$ belong to $B(h)$ and therefore move to the 
left as well.\\ As a consequence, the total time needed to maintain the indices $j(i)$ can be bounded by $\mathcal{O}(\sum_{i=1}^{n} 
m_i)=\mathcal{O}(m)$ since none of these indices is ever decreased.
\paragraph{Representation of cost functions of blocks} In order to realize calls to $PLACE$ and $COLLAPSE$ efficiently, we need some additional data which we store for the blocks consisting of cells we have already processed.
Thereby, the key observation is the fact that in order to implement the function $PLACE$, only local information on the given convex cost function is required since for a convex real function, the question whether the interval where it attains its minimum lies to the left or right of or contains a certain coordinate can be answered by considering local monotonicity properties. In this spirit, for each block $B(i)$, we store the following data: 
\begin{itemize}
 \item a heap $H(i)$ that contains for each $C_l\in B(i)$ the position $p_l^{j(l)+1}-\sum_{h=i}^{l-1}w(C_h)$ unless $j(l)=m_l$ and
 \item the quadratic function $g_i$ defining $f_i$ on the non-empty interval $(\max H(i),x_i]$ (whereby $\max\emptyset:=-\infty$).
\end{itemize}
We outline how to use them in order to implement $PLACE$ and $COLLAPSE$. Consider a call to $PLACE(C_i,\mathcal{L})$ and remember that we implicitly assume that $x_i=x_{max}$ for $1\leq i \leq n$ initially. Further observe that this convention ensures that throughout the algorithm, for $C_h,C_i\in\mathcal{L}$ with $h<i$, we have $x_h+w_h\leq x_i$. In order to execute $PLACE$, the first thing we have to decide is whether $x_h+w_h\leq f_i^+$. While we can compute the value of the left hand side in constant time, $f_i^+$ is not necessarily known to us. However, what we do know is that by convexity of $f_i$, $f_i^+$ is the unique position in $[x_{min}+\sum_{j=1}^{i-1}w(C_j),x_{max}-\sum_{j=i}^n w(C_j)]$ such that $f_i\upharpoonright[x_{min}+\sum_{j=1}^{i-1}w(C_j),x_{max}-\sum_{j=i}^n w(C_j)]$ is monotonically decreasing to its left and strictly monotonically increasing to its right. As a consequence, if $f_i\upharpoonright(\max H(i), x_i]$ (which is given by the quadratic function $g_i$) is monotonically decreasing, we can be sure that $f_i^+\geq x_i\geq x_h+w_h$. On the other hand, as long as $f_i\upharpoonright(\max H(i), x_i]$ is strictly monotonically increasing, we can decrease $x_i$ to $\max\{x_h + w_h, \max H(i)\}$, and, whenever this maximum is attained by $\max H(i)$, pop all corresponding entries from the heap, increment the corresponding indices $j(l)$ by one and insert a new heap entry unless they reach $m_l$, and update $g_i$. Note that if one precomputes all of the values $\sum_{j=1}^{i-1} w(C_j)$, $i=1,\dots,n$ recursively in linear time, which allows to determine $\sum_{j=i}^{l-1} w(C_j)$ in constant time throughout the algorithm, each of these update steps takes constant time per heap entry. In each case where the maximum is not attained by $\max H(i)$, we can infer that $f_i^+<x_h+w_h$ and therefore launch a call of $COLLAPSE$. Finally, if there is some $z\in(\max H(i), x_i)$ where $g_i$ changes from being monotonically decreasing to being strictly monotonically increasing, then $z=f_i^+$ and we are able to decide whether or not $x_h+w_h\leq f_i^+$ holds. In case the latter is true, we also have to determine $\max\{x_h+w_h,f_i^-\}$. To this end, observe that by convexity of $f_i$, $f_i^-$ is the unique coordinate in $[x_{min}+\sum_{j=1}^{i-1}w(C_j),x_{max}-\sum_{j=i}^n w(C_j)]$ such that $f_i$, restricted to the latter interval, is strictly monotonically decreasing to the left, and monotonically increasing to the right of $f_i^-$. By applying a similar strategy as before, we can therefore either compute $f_i^-\in(\max H(i),x_i]$ or set $x_i$ to $\max\{x_h+w_h,\max H(i)\}\geq f_i^-$. As a consequence, we are left with discussing the implementation of $COLLAPSE(C_h,C_i,\mathcal{L})$. Since we do \emph{not} explicitly recompute $f_h^-$ and $f_h^+$ and the updates of $w_h$ and $\mathcal{L}$ can be easily performed in constant time when implementing $\mathcal{L}$ as a doubly linked list, we only have to take care of the redefinition of $f_h$. To this end, note that $g_h$ can be updated by setting $g_h(x)\gets g_h(x)+g_i(x+w_h)$ by a constant number of arithmetic operations on the respective coefficients. As far as the heap $H(h)$ is concerned, we have to shift all entries in $H(i)$ by $w_h$ to the left and then merge $H(i)$ into $H(h)$. By employing \emph{Leftist Heaps} and storing \emph{key differences} instead of the actual keys (see \cite{Tarjan} for further details), the shifting can be performed in constant and the merging in logarithmic (w.r.t.\ the total number of heap elements) time. A logarithmic or even constant time bound also applies for all other heap operations we perform, which comprise the creation of empty heaps, the extraction and deletion of maximum heap entries as well as the insertion of new elements. By observing that the maximum heap size is bounded by $\min\{n,m\}$ since each heap contains at most one entry per cell, but also at most one entry per kink, and that the total number of heap operations is $\mathcal{O}(n+m)$ since for every (pair of) shifting and merging, we remove an entry 
from $\mathcal{L}$, and every kink position is added to and removed from a heap at most once, we obtain the claimed runtime bound.
\section{The Double Row Problem}\label{SecDoubleRowProblem}
In this section, we 
\begin{itemize}
 \item formally introduce the Double Row Problem and
 \item reformulate the feasibility constraints as those of an instance of the Single Row Problem defined on the set of cells of double-row height.
\end{itemize}
 As the name of the problem indicates, the task is to place a set of cells of single- and double-row height within a given rectangular window covering two rows, minimizing a sum of continuous, convex objective functions on the positions of the individual cells. Thereby, the left-to-right ordering of those cells occupying a certain row is fixed and the cells are not allowed to overlap. 
 \begin{definition}[Double Row Problem]
 	\begin{description}
 		\item[]
 		\item[Instance: ]
 		\begin{itemize}
 			\item[]
 			\item a non-empty set $\mathcal{C}:=\{C_1,\dots,C_k\}$ of double-row cells,
 			\item sets of cells \begin{itemize}\item $\mathcal{B}:=\{b_{ij}, i=0,\dots,k, \text{ } j=1,\dots,m_i\}$ and\item $\mathcal{T}:=\{t_{ij}, i=0,\dots,k ,\text{ } j=1,\dots,n_i\}$ \end{itemize}to be placed in the bottom respectively top row,\\ where $m_i,n_i\in\mathbb{N}_0$ for $i=0,\dots,k$, 
 			\item cell widths $w:\mathcal{C}\cup\mathcal{B}\cup\mathcal{T}\rightarrow\mathbb{R}^+$,
 			\item a minimum and  maximum coordinate $x_{min},x_{max}\in\mathbb{R}$ such that 
\[x_{min}+\sum_{i=1}^{k}w(C_i)+\sum_{i=0}^{k}\max\left\{\sum_{j=1}^{m_i}w(b_{ij}),\sum_{j=1}^{n_i}w(t_{ij})\right\}\leq 
x_{max}\] and
 			\item convex, continuous cost functions
 			\begin{itemize}
 				\item $f_i:\mathbb{R}\rightarrow\mathbb{R}$ for $i=1,\dots,k$,
 				\item$g_{ij}:\mathbb{R}\rightarrow \mathbb{R}$ for $i=0,\dots,k$, $j=1,\dots,m_i$ and
 				\item$h_{ij}:\mathbb{R}\rightarrow \mathbb{R}$ for $i=0,\dots,k$, $j=1,\dots,n_i$.
 			\end{itemize}
 		\end{itemize}
 		\item[Task: ] Find coordinates $(x_i)_{i=1}^k$, $(y_{ij}){_{i=0}^k}{_{j=1}^{m_i}}$ and 
$(z_{ij}){_{i=0}^k}{_{j=1}^{n_i}}$ minimizing 
 		\mbox{$\sum_{i=1}^k f_i(x_i) +\sum_{i=0}^k \left(\sum_{j=1}^{m_i} 
g_{ij}(y_{ij})+\sum_{j=1}^{n_i}h_{ij}(z_{ij})\right)$}\\ subject to
 		 \begin{itemize}
 	\item  $x_i+w(C_i)\leq x_{i+1}$ for $i=0,\dots,k$,
 	\item$x_i+w(C_i)\leq y_{i1}$ for $i=0,\dots, k$, 
 	\item$y_{ij}+w(b_{ij})\leq y_{ij+1}$ for $i=0,\dots,k$, $j=1,\dots,m_i-1$,
 	\item $y_{im_i}+w(b_{im_i})\leq x_{i+1}$ for $i=0,\dots, k$,
 	\item$x_i+w(C_i)\leq z_{i1}$ for $i=0,\dots, k$,
 	\item $z_{ij}+w(t_{ij})\leq z_{ij+1}$ for $i=0,\dots,k$, $j=1,\dots,n_i-1$ and 
 	\item $z_{in_i}+w(t_{in_i})\leq x_{i+1}$ for $i=0,\dots, k$,
 \end{itemize}
 	where $x_0:=x_{min}$, $w(C_0):=0$, $x_{k+1}:=x_{max}$ and each constraint only applies if all of its variables exist.
 	\end{description}
 	For $i=0,\dots, k$, we define $\mathcal{B}_i:=\{b_{ij}, j=1,\dots,m_i\}$ and \mbox{$\mathcal{T}_i:=\{t_{ij}, j=1,\dots,n_i\}$}.
 \end{definition}
\begin{figure}
\scalebox{0.7}{
	\begin{tikzpicture}[scale = 0.5, bottom/.style = {draw = blue, fill = blue!50!white}, top/.style = {draw = red!70!black, fill = red!50!white}, double/.style = {draw = green!70!black, fill = {rgb:green,7;black,3;white,10}}]
	\draw[thick] (0,0)--(20,0);
	\draw[thick] (0,1)--(20,1);
	\draw[thick] (0,2)--(20,2);
	\node at (0,-1) {$x_{min}$};
	\node at (20,-1){$x_{max}$};
	\draw[bottom] (1,0) rectangle (4,1);
	\node at (2.5,0.5){\textcolor{blue}{$b_{01}$}};
	\draw[bottom] (8,0) rectangle (9,1);
	\node at (8.5,0.5){\textcolor{blue}{$b_{11}$}};
	\draw[bottom] (9.5,0) rectangle (10.5,1);
	\node at (10,0.5){\textcolor{blue}{$b_{12}$}};
	\draw[bottom] (14.5,0) rectangle (18,1);
	\node at (16.25,0.5){\textcolor{blue}{$b_{21}$}};
	\draw[top] (2,1) rectangle (3,2);
	\node at (2.5,1.5){\textcolor{red!70!black}{$t_{01}$}};
	\draw[top] (7,1) rectangle (9.3,2);
	\node at (8.15,1.5){\textcolor{red!70!black}{$t_{11}$}};
	\draw[top] (10,1) rectangle (11,2);
	\node at (10.5,1.5){\textcolor{red!70!black}{$t_{12}$}};
	\draw[top] (14.5,1) rectangle (15.5,2);
	\node at (15,1.5){\textcolor{red!70!black}{$t_{21}$}};
	\draw[top] (16.5,1) rectangle (17.5,2);
	\node at (17,1.5){\textcolor{red!70!black}{$t_{22}$}};
	\draw[top] (17.5,1) rectangle (18.5,2);
	\node at (18,1.5){\textcolor{red!70!black}{$t_{23}$}};
	\draw[thick, blue, decorate, decoration = {brace, amplitude = 5pt}] (5,0)--(0,0);
	\node at (2.5,-1) {\textcolor{blue}{$\mathcal{B}_0$}};
	\draw[thick, blue, decorate, decoration = {brace, amplitude = 5pt}] (11.5,0)--(7,0);
	\node at (9.25,-1) {\textcolor{blue}{$\mathcal{B}_1$}};
	\draw[thick, blue, decorate, decoration = {brace, amplitude = 5pt}] (20,0)--(14,0);
	\node at (17,-1) {\textcolor{blue}{$\mathcal{B}_2$}};
	\draw[thick, red!70!black, decorate, decoration = {brace, amplitude = 5pt}] (0,2)--(5,2);
	\node at (2.5,3) {\textcolor{red!70!black}{$\mathcal{T}_0$}};
	\draw[thick, red!70!black, decorate, decoration = {brace, amplitude = 5pt}] (7,2)--(11.5,2);
	\node at (9.25,3) {\textcolor{red!70!black}{$\mathcal{T}_1$}};
	\draw[thick, red!70!black, decorate, decoration = {brace, amplitude = 5pt}] (14,2)--(20,2);
	\node at (17,3) {\textcolor{red!70!black}{$\mathcal{T}_2$}};
	\draw[double] (5,0) rectangle (7,2);
	\node at (6,1){\textcolor{green!30!black}{$C_1$}};
	\draw[double] (11.5,0) rectangle (14,2);
	\node at (12.75,1){\textcolor{green!30!black}{$C_2$}};
	\end{tikzpicture}}
	\Description{The figure depicts two horizontal rows comprising several rectangular cells of single as well as double-row height that are aligned with the row boundaries. The boundaries of the $x$-interval spanned by the two rows are named $x_{min}$ and $x_{max}$, respectively. The two cells of double-row height that are present are labeled $C_1$ and $C_2$ (from left to right). The gaps to the left of $C_1$, between $C_1$ and $C_2$ and to the right of $C_2$ contain cells of single-row height, for which the naming scheme works as follows: Single row cells in the bottom row are named $b$, while those in the top row are called $t$. Each cell name further comprises two indices the first one of which indicates the gap it is placed in (ranging from $0$ to $k$ from left to right), while the second one reflects the left-to-right ordering within gap and row. For example, the two bottom row cells in the gap between $C_1$ and $C_2$ are labeled $b_{11}$ and $b_{12}$.}
	\caption{The Double Row Problem.}
\end{figure}
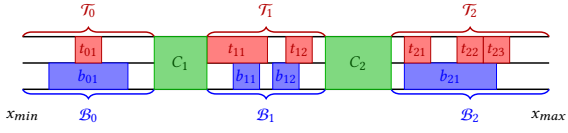
 \begin{proposition}
 	Given a tuple $(x^*_i)_{i=1}^k$ and an instance of the Double Row Problem as defined above, there exists a feasible solution to the Double Row Problem with $x_i=x^*_i$ for  $i=1,\dots,k$ if and only if $$x^*_i+w(C_i)+\max\left\{\sum_{j=1}^{m_i} w(b_{ij}), \sum_{j=1}^{n_i}w(t_{ij})\right\}\leq x^*_{i+1}\text{ for } i=0,\dots,k,$$ where $x^*_0:=x_0:=x_{min}$, $w(C_0):=0$ and $x^*_{k+1}:=x_{k+1}:=x_{max}$.\label{PropFeasible}
 \end{proposition}
We call such a tuple $(x^*_i)_{i=1}^k$ \emph{feasible}.
 \begin{remark}
  Note that a tuple $(x^*_i)_{i=1}^k$ is feasible if and only if it defines a feasible solution to the instance of the Single Row Problem with cell set $\mathcal{C}$, cell widths \[w'(C_i):=w(C_i)+\max\left\{\sum_{j=1}^{m_i} w(b_{ij}), \sum_{j=1}^{n_i}w(t_{ij})\right\}\] and enclosing $x$-interval $[x'_{min},x'_{max}]$ given by \[x'_{min}:=x_{min}+\max\left\{\sum_{j=1}^{m_0} w(b_{0j}), \sum_{j=1}^{n_0}w(t_{0j})\right\}\] and $x'_{max}:=x_{max}$.
 \end{remark}

\section{Reduction to the Single Row Problem}\label{SecReduction}
For the remainder of this paper, we restrict ourselves to the case of piecewise quadratic cost functions and show how to reduce the respective variant of the Double Row Problem to the Single Row one.
As we have already seen how to deal with the subject of feasibility, it remains to transfer costs from the single-row cells to the double-row ones, i.e.\ to determine the minimum cost of a feasible extension of a feasible tuple $(x^*_i)_{i=1}^k$ and to express it as $\sum_{i=1}^k f'_i(x^*_i)$ for some piecewise quadratic objective functions $f'_i$.
\begin{itemize}
 \item We examine the structure of an optimum extension of a feasible tuple to coordinates for the single-row height cells.
 \item Lemma~\ref{LemNewCostFuns} expresses the total cost of such an extension, up to a constant, as a sum $\sum_{i=1}^k F_i(x^*_i)$.
 \item We show that each of the functions $F_i$ is convex and piecewise quadratic and linearly bound the total number of kinks.
 \item We then derive our main result stated in Theorem~\ref{TheoMainTheorem}.
\end{itemize}
Consider the coordinates $(\bar{y}_{ij}){_{i=0}^k}{_{j=1}^{m_i}}$ and $(\bar{z}_{ij}){_{i=0}^k}{_{j=1}^{n_i}}$ arising from runs of the Clumping Algorithm on the instances of the Single Row Problem given by \mbox{$(\mathcal{B}_i,w\upharpoonright\mathcal{B}_i,x_{min},x_{max},(g_{ij})_{j=1}^{m_i})$} and $(\mathcal{T}_i,$ $w\upharpoonright\mathcal{T}_i,x_{min},x_{max},(h_{ij})_{j=1}^{n_i})$ for $i=0,\dots, k$. Note that once a feasible tuple $(x^*_i)_{i=1}^k$ of coordinates for the double-row cells has been fixed, coordinates $(y_{ij}){_{i=0}^k}{_{j=1}^{m_i}}$ and $(z_{ij}){_{i=0}^k}{_{j=1}^{n_i}}$ extend them to a feasible solution of the Double Row Problem if and only if for each $i\in\{0,\dots,k\}$, $(y_{ij})_{j=1}^{m_i}$ and $(z_{ij})_{j=1}^{n_i}$ constitute feasible solutions of the instances of the Single Row Problem given by \mbox{$(\mathcal{B}_i,w\upharpoonright\mathcal{B}_i,x^*_i+w(C_i),x^*_{i+1},(g_{ij})_{j=1}^{m_i})$} and $(\mathcal{T}_i,w\upharpoonright\mathcal{T}_i,x^*_i+w(C_i),$ $x^*_{i+1},(h_{ij})_{j=1}^{n_i})$, respectively, whereby again $x^*_0:=x_{min}$, \mbox{$w(C_0):=0$} and $x^*_{k+1}:=x_{max}$. Note that these instances are feasible by feasibility of $(x^*_i)_{i=1}^k$. But now, since for each $i=0,\dots,k$, we have $x_{min}\leq x^*_i+w(C_i)\leq x^*_{i+1}\leq x_{max}$, Theorem~\ref{TheoOptSolutionShrinkInterval} tells us that an optimum extension $(y^*_{ij}){_{i=0}^k}{_{j=1}^{m_i}}$ and $(z^*_{ij}){_{i=0}^k}{_{j=1}^{n_i}}$ of $(x^*_i)_{i=1}^k$ is given by 
\begin{equation}y^*_{ij}=\min\{x^*_{i+1}-\sum_{l=j}^{m_i} w(b_{il}),\max\{x^*_i+w(C_i)+\sum_{l=1}^{j-1} w(b_{il}),\bar{y}_{ij}\}\}\label{Eqystar}\end{equation} and
\begin{equation}z^*_{ij}=\min\{x^*_{i+1}-\sum_{l=j}^{n_i} w(t_{il}),\max\{x^*_i+w(C_i)+\sum_{l=1}^{j-1} w(t_{il}),\bar{z}_{ij}\}\}.\label{Eqzstar}\end{equation}
This allows us to express the total cost of the solution in terms of the coordinates $(x^*_i)_{i=1}^k$:
\begin{lemma}
 Let $(\bar{y}_{ij}){_{i=0}^k}{_{j=1}^{m_i}}$ and $(\bar{z}_{ij}){_{i=0}^k}{_{j=1}^{n_i}}$ as before and define
 \begin{align}F_i:x\mapsto &f_i(x) \label{SummandF}\\ +& \sum_{j=1}^{m_{i-1}} g_{i-1j}(\min\{x-\sum_{l=j}^{m_{i-1}}w(b_{i-1l}), \bar{y}_{i-1j}\})\label{SummandGmin}\\ +&\sum_{j=1}^{m_i}g_{ij}(\max\{x+w(C_i)+\sum_{l=1}^{j-1}w(b_{il}),\bar{y}_{ij}\})\label{SummandGmax}\\
 	+&\sum_{j=1}^{n_{i-1}} h_{i-1j}(\min\{x-\sum_{l=j}^{n_{i-1}}w(t_{i-1l}), \bar{z}_{i-1j}\}) \label{SummandHmin}\\+&\sum_{j=1}^{n_i}h_{ij}(\max\{x+w(C_i)+\sum_{l=1}^{j-1}w(t_{il}),\bar{z}_{ij}\})\label{SummandHmax}\end{align} and 
 	$c:=\sum_{i=1}^{k-1}\sum_{j=1}^{m_i} g_{ij}(\bar{y}_{ij}) +\sum_{i=1}^{k-1}\sum_{j=1}^{n_i} h_{ij}(\bar{z}_{ij})$. Then for a feasible tuple $(x^*_i)_{i=1}^k$, the total cost of an optimum solution to the Double Row Problem with $x_i=x^*_i$ for $i=1,\dots,k$ amounts to $\sum_{i=1}^k F_i(x^*_i)-c$.\label{LemNewCostFuns}
\end{lemma}
\begin{proof}
	Recall that an optimum extension $(y^*_{ij}){_{i=0}^k}{_{j=1}^{m_i}}$ and $(z^*_{ij}){_{i=0}^k}{_{j=1}^{n_i}}$ of $(x^*_i)_{i=1}^k$ is given by
	\eqref{Eqystar} and \eqref{Eqzstar}.
We are done if we can show that for any cell, the part of the cost term involving its objective function matches the cost of its position in the given solution.\\
 	For the cells $(C_i)_{i=1}^k$, this is clear.\\
 	For a cell $b_{0j}$ with $j\in\{1,\dots,m_0\}$, the desired statement follows from $x^*_0+w(C_0)+\sum_{l=1}^{j-1} w(b_{0l})=x_{min}+\sum_{l=1}^{j-1} w(b_{0l})\leq \bar{y}_{0j},$ and a similar argument applies for $i=k$.\\
 	For a cell $b_{ij}$ with $i\in\{1,\dots,k-1\}$ and $j\in\{1,\dots,m_i\}$, we exemplarily consider the case where $\bar{y}_{ij}\leq x^*_i+w(C_i)+\sum_{l=1}^{j-1}w(b_{il})$ since the cases $x^*_i+w(C_i)+\sum_{l=1}^{j-1}w(b_{il})< \bar{y}_{ij} < x^*_{i+1}-\sum_{l=j}^{m_{i}}w(b_{il})$ and $x^*_{i+1}-\sum_{l=j}^{m_{i}}w(b_{il})\leq \bar{y}_{ij}$ can be treated similarly. In the mentioned case, we get
 	\begin{align*}y^*_{ij}&=\min\{x^*_{i+1}-\sum_{l=j}^{m_i}w(b_{il}),\max\{x^*_i+w(C_i)+\sum_{l=1}^{j-1}w(b_{il}), \bar{y}_{ij}\}\}\\
 		&=\max\{x^*_i+w(C_i)+\sum_{l=1}^{j-1}w(b_{il}), \bar{y}_{ij}\} \end{align*} and $\min\{x^*_{i+1}-\sum_{l=j}^{m_{i}}w(b_{il}), \bar{y}_{ij}\}=\bar{y}_{ij},$
 		so \begin{align*}
 		&g_{ij}(\max\{x^*_i+w(C_i)+\sum_{l=1}^{j-1}w(b_{il}),\bar{y}_{ij}\})\\ &+  g_{ij}(\min\{x^*_{i+1}-\sum_{l=j}^{m_{i}}w(b_{il}), \bar{y}_{ij}\})-g_{ij}(\bar{y}_{ij}) \\
 		=& g_{ij}(y^*_{ij})+g_{ij}(\bar{y}_{ij})-g_{ij}(\bar{y}_{ij})=g_{ij}(y^*_{ij}).
 		\end{align*}
 	The cells in $\mathcal{T}$ can be treated analogously.\end{proof}
 	Up to the constant $c$, which only depends on the given instance of the Double Row Problem, but not on the tuple $(x^*_i)_{i=1}^k$, we can hence express the costs of an optimum solution extending a feasible tuple $(x^*_i)_{i=1}^k$ as a sum of the cost functions $(F_i)_{i=1}^k$ applied to the individual coordinates. Note that each of the summands contributing to $F_i$ and hence $F_i$ itself is piecewise quadratic since linear shifting as well as replacement by a constant function to the left or right of a certain coordinate (ensuring continuity) preserves this property. In addition to that, it is not hard to see that the total number of kinks the cost functions $(F_i)_{i=1}^k$ possess can be bounded by $2\cdot(|\mathcal{B}|+|\mathcal{T}|)+N$, where $N$ denotes the total number of kinks present in the cost functions of the single- and double-row cells. To show that all $F_i$ are actually convex, it is sufficient to show that each of the summands \eqref{SummandF}-\eqref{SummandHmax} induces a convex function. This is clear for \eqref{SummandF}, and we exemplarily show it for \eqref{SummandGmax}. Let $\mathcal{L}_i^b$ denote the list of cells arising from the run of the Clumping Algorithm on the aforementioned instance of the Single Row Problem with cell set $\mathcal{B}_i$. Given that for $b_{ij}\in\mathcal{L}_i^b$, the cells in the block $B(ij)$ starting at $b_{ij}$ are placed contiguously, we can rewrite \eqref{SummandGmax} as  \mbox{$\sum_{b_{ij}\in \mathcal{L}_i^b}G_{ij}(\max\{x+w(C_i)+\sum_{l=1}^{j-1}w(b_{il}),\bar{y}_{ij}\})$}, where $G_{ij}$ denotes the cumulated cost function of the block represented by $b_{ij}$. Recall that by definition of the Clumping Algorithm, $\bar{y}_{ij}$ occupies a minimum position of $G_{ij}$ for $b_{ij}\in\mathcal{L}_i^b$. Given that for a continuous, convex function $f:[a,b]\rightarrow\mathbb{R}$ and \mbox{$x_0\in\mathrm{argmin}\{f(x),x\in[a,b]\}$}, the function mapping $x\in[a,b]$ to $f(\max\{x,x_0\})$ is convex, it follows that \eqref{SummandGmax} defines a convex function in $x$. By applying analogous arguments for the remaining summands, we can infer that each $F_i$ is convex as a sum of convex functions. This completes our reduction from the Double to the Single Row Problem and it remains to discuss the runtime it requires. Note that the positions $(\bar{y}_{ij}){_{i=0}^k}{_{j=1}^{m_i}}$ and $(\bar{z}_{ij}){_{i=0}^k}{_{j=1}^{n_i}}$ can be computed in total time
 	$\mathcal{O}((|\mathcal{B}|+|\mathcal{T}|+N)\cdot\log(|\mathcal{B}|+|\mathcal{T}|))$, where again $N$ denotes the total number of kinks of the all cost functions appearing in the given instance of the Double Row Problem.\\ A time of \mbox{$\mathcal{O}((|\mathcal{C}|+|\mathcal{B}|+|\mathcal{T}|+N)\cdot\log(|\mathcal{C}|+|\mathcal{B}|+|\mathcal{T}|+N))$} then suffices to build up and solve the instance of the Single Row Problem on the set of double-row cells to which we reduce, and optimum coordinates for the single-row cells can be deduced from the computed positions for the cells in $\mathcal{C}$ in linear time. Putting everything together, we can therefore formulate the following theorem:
 \begin{theorem}
 	The Double Row Problem with piecewise quadratic functions with a total amount of $N$ kinks can be solved in time $\mathcal{O}((|\mathcal{C}|+|\mathcal{B}|+|\mathcal{T}|+N)\cdot \log(|\mathcal{C}|+|\mathcal{B}|+|\mathcal{T}|+N)).$\label{TheoMainTheorem}
 \end{theorem}
\section{Experimental Results}\label{SecPracticalResults}
\begin{figure}[t]
	\centering
	\includegraphics[trim = 0cm 1cm 0cm 1cm , clip, width=0.6\columnwidth]{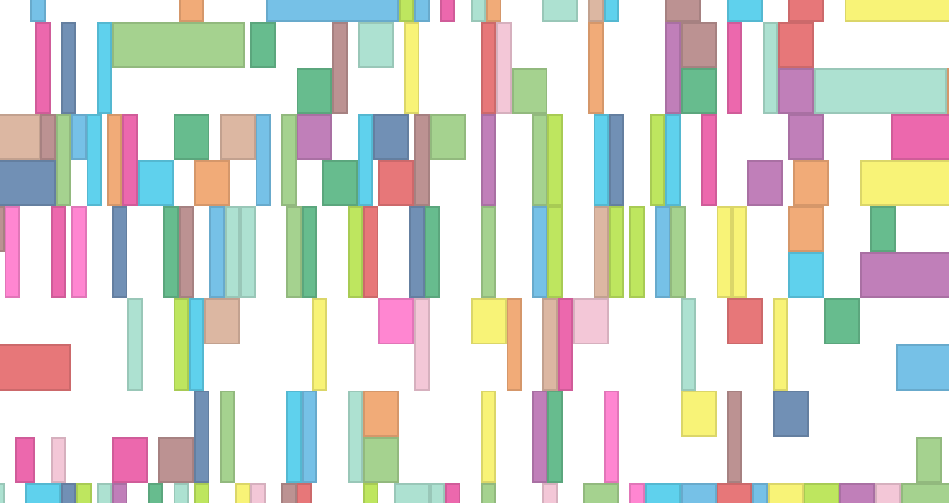}
	\captionof{figure}{superblue12}\label{Figsuperblue}
	\Description{The image shows an excerpt of a legalized placement on the chip superblue12, where the cells of double 
		row height are quite spaced out.}
\end{figure}

\begin{figure}[t]
	\centering
	\includegraphics[width=0.6\columnwidth]{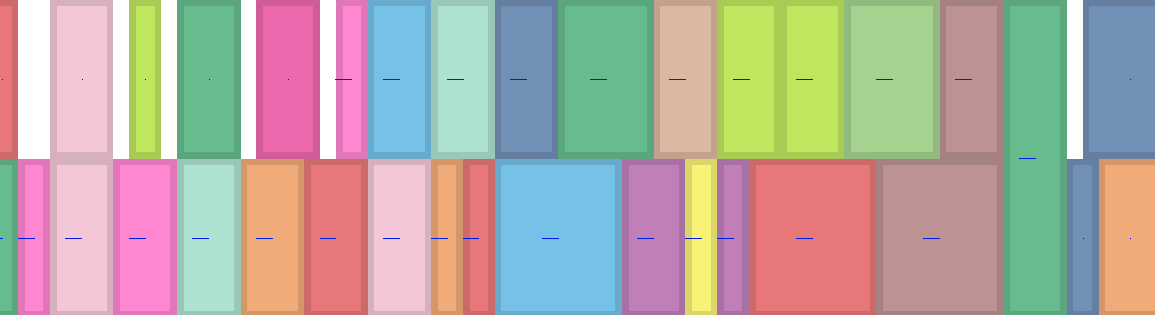}
	\Description{The image shows an excerpt of two rows of a legalized placement on the chip matrix\_mult\_1, where a large block of consecutive cells of single-row height and the neighboring cell of double-row height shift to the right once the latter is unfixed.}
	\captionof{figure}{matrix\_mult\_1 after our algorithm. Blue lines indicate movement w.r.t.\ the output of 
		TCAD'13.}\label{Figmatrixmult}
\end{figure}

 \begin{table*}[t]
 \caption{Comparison between the average cell movement in terms of horizontal placement sites.}\label{TableAvMovement}
 \begin{adjustbox}{width=0.9\textwidth}
  \begin{tabular}{|l|r|rrrr|rrrrr|rrrr|rrr|}
  \toprule
  Instance & GP HPWL (m)& \multicolumn{4}{c|}{$\Delta$ HPWL} & \multicolumn{5}{c|}{Av. L1 Movement (Sites)} & \multicolumn{4}{c|}{Max. L1 Movement (Sites)} & \multicolumn{3}{c|}{CPU (sec)} \\
  & & DAC'17 & ISPD'19 & TCAD'13 & Ours & DAC'17 & ISPD'19 & TCAD'13 & Ours & $\dfrac{\text{Ours}}{\text{ISPD'19}}$ & DAC'17 & ISPD'19 & TCAD'13 & Ours & DAC'17 & ISPD'19  & Ours\\
  \midrule
  des\_perf\_1	& 1.217	& 16.21\%	& 6.66\%	& 4.52\%	& 4.52\%	& 10.86	& 6.97	& 6.66	& 6.66	& 95.55\%	& 200.82	& 48.95	& 57.22	& 57.22	& 11.23	& 11.75	& 9.97 \\
des\_perf\_ a\_md1	& 2.160	& 3.27\%	& 2.48\%	& 2.20\%	& 2.19\%	& 6.71	& 5.94	& 5.85	& 5.79	& 97.47\%	& 607.30	& 607.30	& 607.30	& 607.30	& 2.30	& 2.79	& 8.05\\
des\_perf\_a\_md2	& 2.177	& 3.35\%	& 2.51\%	& 2.23\%	& 2.23\%	& 6.77	& 5.93	& 6.08	& 6.07	& 102.36\%	& 403.86	& 403.86	& 403.86	& 403.86	& 2.19	& 6.82	& 8.53\\
des\_perf\_b\_md1	& 2.106	& 1.75\%	& 1.52\%	& 1.61\%	& 1.59\%	& 5.17	& 4.77	& 4.78	& 4.72	& 98.95\%	& 79.34	& 38.45	& 48.19	& 45.19	& 2.01	& 3.64	& 6.79\\
des\_perf\_b\_md2	& 2.137	& 2.05\%	& 1.72\%	& 1.50\%	& 1.49\%	& 5.74	& 5.25	& 5.38	& 5.31	& 101.14\%	& 198.74	& 39.76	& 50.68	& 50.68	& 2.31	& 3.12	& 8.06\\
edit\_dist\_1\_md1	& 4.004	& 1.47\%	& 1.39\%	& 1.27\%	& 1.26\%	& 6.22	& 5.79	& 5.75	& 5.69	& 98.27\%	& 109.34	& 95.45	& 67.55	& 67.55	& 3.49	& 5.19	& 9.67\\
edit\_dist\_a\_md2	& 5.103	& 1.17\%	& 1.01\%	& 0.92\%	& 0.91\%	& 6.02	& 5.51	& 5.57	& 5.51	& 100.00\%	& 164.00	& 164.00	& 164.00	& 164.00	& 2.59	& 2.24	& 10.78\\
edit\_dis\_ a\_md3	& 5.328	& 2.69\%	& 1.48\%	& 1.02\%	& 1.02\%	& 9.11	& 7.08	& 6.96	& 6.93	& 97.88\%	& 233.00	& 233.00	& 233.00	& 233.00	& 5.91	& 15.68	& 15.87\\
fft\_2\_md2	& 0.444	& 11.21\%	& 8.78\%	& 7.14\%	& 7.02\%	& 8.84	& 7.54	& 7.89	& 7.76	& 102.92\%	& 102.94	& 73.60	& 59.55	& 60.55	& 0.70	& 2.89	& 2.81\\
fft\_a\_md2	& 1.092	& 0.98\%	& 0.95\%	& 1.13\%	& 1.13\%	& 5.03	& 4.86	& 4.74	& 4.70	& 96.71\%	& 345.50	& 345.50	& 343.48	& 346.50	& 0.69	& 0.60	& 2.15\\
ff\_ a\_md3	& 0.949	& 1.08\%	& 1.08\%	& 1.22\%	& 1.22\%	& 4.73	& 4.55	& 4.43	& 4.42	& 97.14\%	& 109.62	& 109.62	& 102.59	& 102.59	& 0.63	& 0.40	& 1.91\\
pci\_bridge32\_a\_md1	& 0.454	& 3.61\%	& 3.38\%	& 3.00\%	& 2.95\%	& 6.01	& 5.64	& 5.83	& 5.76	& 102.13\%	& 72.48	& 63.76	& 63.76	& 63.76	& 0.61	& 2.29	& 2.01\\
pci\_bridge32\_a\_md2	& 0.565	& 8.33\%	& 4.38\%	& 3.68\%	& 3.62\%	& 9.43	& 7.14	& 7.55	& 7.45	& 104.34\%	& 186.08	& 121.35	& 121.35	& 121.35	& 0.53	& 3.34	& 3.76\\
pc\_ bridge32\_b\_md1	& 0.660	& 2.55\%	& 2.26\%	& 2.13\%	& 2.11\%	& 6.35	& 6.01	& 5.79	& 5.72	& 95.17\%	& 322.71	& 332.71	& 313.99	& 313.99	& 0.52	& 0.70	& 2.41\\
pci\_bridge32\_b\_md2	& 0.574	& 2.80\%	& 2.53\%	& 2.57\%	& 2.57\%	& 5.92	& 5.53	& 5.43	& 5.42	& 98.01\%	& 640.12	& 430.04	& 430.04	& 430.04	& 0.50	& 0.66	& 1.89\\
pci\_bridge32\_b\_md3	&0.583	& 3.63\%	& 3.17\%	& 3.14\%	& 3.13\%	& 6.74	& 6.10	& 6.13	& 6.12	& 100.33\%	& 398.57	& 398.57	& 398.58	& 398.58	& 0.51	& 1.58	& 2.21\\\midrule
average	& 	& 4.13\%	& 2.83\%	& 2.46\%	& 2.44\%	& 6.85	& 5.91	& 5.93	& 5.88	& 99.27\%	& 260.90	& 219.12	& 216.57	& 216.64	& 2.30	& 3.98	&5.06\\\bottomrule
\end{tabular}
 \end{adjustbox}
 \end{table*}
 	\begin{table}[t]
 	 \caption{Comparison between the squared cell movement resulting from the legalization algorithm described in TCAD'13 and our algorithm.} \label{TableQuadMovement}
  \begin{adjustbox}{width=\columnwidth}
 \begin{tabular}{|l|r|rrr|rrr|}
\toprule
Instance & GP HPWL (m) & \multicolumn{3}{c|}{ Cells} & \multicolumn{3}{c|}{Squared Cell Movement}\\ 
 &  &  Single & Double & $\dfrac{\text{\# double}}{\text{\# cells}}$ & TCAD'13 &  Ours & $\dfrac{\text{Ours}}{\text{TCAD'13}}$ \\ \midrule
 
 des\_perf\_1	& 1.433	& 103842	& 8802	& 7.81\%	& 4.15E+10	& 2.82E+10	& 68.00\%\\
des\_perf\_a	& 2.573	& 99775	& 8513	& 7.86\%	& 3.66E+09	& 2.51E+09	& 68.53\%\\
des\_perf\_b	& 2.131	& 103842	& 8802	& 7.81\%	& 3.84E+09	& 2.49E+09	& 64.94\%\\
edit\_dist\_a	& 5.252	& 121913	& 5500	& 4.32\%	& 4.49E+09	& 3.17E+09	& 70.55\%\\
fft\_1	& 0.456	& 30297	& 1984	& 6.15\%	& 9.53E+09	& 5.54E+09	& 58.18\%\\
fft\_2	& 0.463	& 30297	& 1984	& 6.15\%	& 1.95E+09	& 1.20E+09	& 61.43\%\\
fft\_a	& 0.750	& 28718	& 1907	& 6.23\%	& 1.30E+09	& 9.04E+08	& 69.41\%\\
fft\_b	& 0.952	& 28718	& 1907	& 6.23\%	& 1.89E+09	& 1.27E+09	& 67.13\%\\
matrix\_mult\_1	& 2.391	& 152427	& 2898	& 1.87\%	& 9.80E+09	& 6.81E+09	& 69.47\%\\
matrix\_mult\_2	& 2.584	& 152427	& 2898	& 1.87\%	& 8.26E+09	& 5.68E+09	& 68.77\%\\
matrix\_mult\_a	& 3.772	& 146837	& 2813	& 1.88\%	& 2.97E+09	& 2.31E+09	& 77.89\%\\
matrix\_mult\_b	& 3.299	& 143695	& 2740	& 1.87\%	& 2.61E+09	& 2.15E+09	& 82.07\%\\
pci\_bridge32\_a	& 0.460	& 26268	& 3249	& 11.01\%	& 1.23E+09	& 7.93E+08	& 64.60\%\\
pci\_bridge32\_b	& 0.980	& 25734	& 3180	& 11.00\%	& 6.13E+08	& 3.61E+08	& 58.86\%\\
superblue11\_a	& 42.915	& 861314	& 64302	& 6.95\%	& 2.67E+11	& 2.48E+11	& 92.74\%\\
superblue12	& 39.110	& 1172586	& 114362	& 8.89\%	& 5.61E+11	& 5.38E+11	& 95.84\%\\
superblue14	& 27.905	& 564769	& 47474	& 7.75\%	& 2.00E+11	& 1.81E+11	& 90.19\%\\
superblue16 a	& 31.330	& 625419	& 55031	& 8.09\%	& 6.35E+10	& 4.76E+10	& 74.99\%\\
superblue19	& 20.722	& 478109	& 27988	& 5.53\%	& 1.22E+11	& 1.14E+11	& 93.21\%\\\midrule
average & & & & & & & 73.51\%\\\bottomrule
 \end{tabular}
 \end{adjustbox}
\end{table}
 We implemented the proposed algorithm in the C++ programming language and embedded it into the legalization framework described in \cite{BrennerMCF}. More precisely, we first run the legalization algorithm from \cite{BrennerMCF}, which legalizes all cells of more than single-row height via a greedy projection approach and then proceeds by assigning all cells of single-row height to so-called \emph{zones}, unblocked segments of cell rows, through a min-cost-flow algorithm. Within each zone, the left-to-right ordering is inferred from the Global Placement positions. While the algorithm from \cite{BrennerMCF} proceeds by optimizing squared cell movement only within each zone making use of the Clumping Algorithm, we instead apply the Double Row Algorithm to the instances of the Double Row Problem arising from the given left-to-right ordering in every second pair of rows, treating all cells of more than double-row height as blockages.\\ All experiments were performed single-threaded on Intel Xeon 3.3GHz CPUs with 384GB RAM. We conduct two experiments on two different sets of benchmarks. The first one aims at establishing the competitiveness of our legalization approach when compared to recent works on the matter of mixed-cell-height legalization. The second experiment displays the effectiveness of the Double Row Algorithm in improving squared cell movement.\\ For the first experiment, we run our algorithm on benchmark instances from the ICCAD-2017 CAD Contest on Multi-Deck Standard-Cell Legalization \cite{benchmarks}. In doing so, we omit fence region constraints as well as soft constraints, but stick to the required power-rail alignment. As most prior works optimize linear instead of squared cell movement, we employ our proposed legalization method to minimize linear movement during the Double Row Algorithm. Observe that this is possible since for each cell, once its row assignment is fixed, the distance to its Global Placement location constitutes a piecewise linear and hence in particular piecewise quadratic function. However, we point out that minimizing l1 movement is not the main purpose of our algorithm and that in particular, the assignment to zones is designed to optimize squared instead of linear movement. Hence, the subsequent comparison should be regarded as proof that our algorithm, even though not explicitly devised to do so, can compete with state-of-the-art legalizers concerning linear cell movement. We compare the average l1 cell movement achieved by our algorithm to the results obtained by \cite{QPLegalization2017} and the state-of-the-art paper \cite{Analytical} as reported in \cite{Analytical} as well as the legalization approach from \cite{BrennerMCF}. Table~\ref{TableAvMovement} displays the relative increase ($\Delta$~HPWL) of the half-perimeter wire length after Global Placement (GP HPWL), the average l1 cell movement (measured in horizontal placement sites), the maximum l1 cell movement (again measured in placement sites) and the runtime in CPU seconds for the algorithms in \cite{QPLegalization2017}(DAC'17), \cite{Analytical}(ISPD'19) and \cite{BrennerMCF}(TCAD '13) and the algorithm suggested in this paper (Ours). Concerning the average cell movement, which we are mainly interested in for this comparison, the column labeled ''Ours/ISPD'19'' contains the percentages the average cell movement obtained by "Ours" constitutes of the average cell movement reported by ISPD'19 \cite{Analytical}. The final row labeled ''average'' displays the average of all prior values in the respective column. In particular, the respective entry in the column ''Ours/ISPD'19'' refers to the average of the above percentages. One can see that on average, our proposed algorithm achieves comparable results to the algorithm in \cite{Analytical}, which in turn produces considerably better results than \cite{QPLegalization2017} when it comes to average cell movement. However, the deviation between the 
different instances is relatively high: While there are some on which our algorithm significantly outperforms the method from \cite{Analytical} (including those where no cells of triple- and quadruple-row height are present), the converse is true for several other test cases. One possible explanation for this might be the fact that the greedy legalization of cells of more than double-row height only works well if they are sufficiently spaced out in the Global Placement solution, which is true for only some of the given benchmarks. When it comes to running time, maximum movement, and increase in HPWL, our algorithm can be seen to yield comparable or even better results.\\ In our second experiment, we compare the total quadratic cell movement achieved by the algorithm described in \cite{BrennerMCF} to minimize squared cell movement and our new method. As the number of double-row cells on the ICCAD-2017 CAD Contest benchmarks \cite{benchmarks} is rather small, we employ a set of benchmarks generated by the authors of \cite{ChowPuiYoung} by modifying instances from the ISPD 2015 Detailed Routing-Driven Placement Contest \cite{bustany2015ISPD}. While these are more suitable for the primary application of our algorithm, we decided against using them for a comparison to other legalizers since they are not publicly available and the parsing process appears to be more error-prone due to a non-standard format. For completeness, we nevertheless state that our experiments revealed an average cell movement better than the one obtained by \cite{ChowPuiYoung}, \cite{QPLegalization2017} and \cite{EffectiveLegalizationAlgorithm}, but worse than what is claimed in \cite{hung2017mixed} (at the cost of a considerably higher runtime) and \cite{Routability}.\\The results of our second experiment can be read from Table~\ref{TableQuadMovement}, which displays the squared cell movement achieved by the algorithm described in TCAD'13 \cite{BrennerMCF} and the algorithm proposed in this paper. The first column contains the instance name, while the columns labeled "Single" and "Double" display the number of cells of single- respectively double-row height present on the given test case, whereby the fraction the number of double-row cells constitutes of the total number of cells can be found in the following column labeled "\# double/\# cells". Cells of more than double-row height do not occur. The last three columns contain the total quadratic cell movement in squared base units resulting from the TCAD'13 legalization algorithm and ours as well as the ratio between both. One can see that an average percental decrease in quadratic movement of more than $26\%$ is achieved. Even on instances with only a few cells of double-row height, improvements achieved by the application of the Double Row Algorithm are quite significant, which can be explained by the fact that even a single double-row cell being fixed in position may lead to the displacement of huge blocks of consecutive cells of single-row height in densely packed regions (see Figure~\ref{Figmatrixmult}). On the other hand, if many of the cells of double-row height do not interfere with those of single-row height at all in that there is sufficient horizontal whitespace around them, comparably small improvements are obtained despite a considerable number of cells of double-row height present (see Figure~\ref{Figsuperblue}). However, as the legalization task becomes more difficult in those cases where the Global Placement packs the cells relatively dense locally, the Double Row Algorithm can be considered a worthwhile extension of the considered legalization framework.
\section{Conclusion}
In this paper, we have presented a fast algorithm to minimize quadratic (or linear) cell displacement for pairs of cell rows comprising cells of both single- and double-row height with predefined target locations and a fixed left-to-right ordering. Even though the surrounding legalization framework is designed to optimize squared instead of linear cell displacement, our results are competitive when compared to state-of-the-art works on mixed-cell-height legalization. Moreover, experimental results comparing the squared cell displacement when fixing all cells of double-row height and when employing the Double Row Algorithm, respectively, clearly speak in favor of its effectiveness. 

\bibliographystyle{ACM-Reference-Format}
\bibliography{legalization_paper.bib}


\begin{thebibliography}{25}


\ifx \showCODEN    \undefined \def \showCODEN     #1{\unskip}     \fi
\ifx \showDOI      \undefined \def \showDOI       #1{#1}\fi
\ifx \showISBNx    \undefined \def \showISBNx     #1{\unskip}     \fi
\ifx \showISBNxiii \undefined \def \showISBNxiii  #1{\unskip}     \fi
\ifx \showISSN     \undefined \def \showISSN      #1{\unskip}     \fi
\ifx \showLCCN     \undefined \def \showLCCN      #1{\unskip}     \fi
\ifx \shownote     \undefined \def \shownote      #1{#1}          \fi
\ifx \showarticletitle \undefined \def \showarticletitle #1{#1}   \fi
\ifx \showURL      \undefined \def \showURL       {\relax}        \fi
\providecommand\bibfield[2]{#2}
\providecommand\bibinfo[2]{#2}
\providecommand\natexlab[1]{#1}
\providecommand\showeprint[2][]{arXiv:#2}

\bibitem[\protect\citeauthoryear{Brenner}{Brenner}{2013}]%
        {BrennerMCF}
\bibfield{author}{\bibinfo{person}{U. Brenner}.}
  \bibinfo{year}{2013}\natexlab{}.
\newblock \showarticletitle{Bonn{P}lace {L}egalization: {M}inimizing {M}ovement
  by {I}terative {A}ugmentation}.
\newblock \bibinfo{journal}{\emph{TCAD}} \bibinfo{volume}{32},
  \bibinfo{number}{8} (\bibinfo{year}{2013}), \bibinfo{pages}{1215--1227}.
\newblock


\bibitem[\protect\citeauthoryear{{Brenner} and {Vygen}}{{Brenner} and
  {Vygen}}{2000}]%
        {BrennerVygen}
\bibfield{author}{\bibinfo{person}{U. {Brenner}} {and} \bibinfo{person}{J.
  {Vygen}}.} \bibinfo{year}{2000}\natexlab{}.
\newblock \showarticletitle{{F}aster {O}ptimal {S}ingle-{R}ow {P}lacement with
  {F}ixed {O}rdering}. In \bibinfo{booktitle}{\emph{Proceedings Design,
  Automation and Test in Europe}}. \bibinfo{pages}{117--121}.
\newblock


\bibitem[\protect\citeauthoryear{Brenner and Vygen}{Brenner and Vygen}{2004}]%
        {LegalizationBonnTools}
\bibfield{author}{\bibinfo{person}{U. Brenner} {and} \bibinfo{person}{J.
  Vygen}.} \bibinfo{year}{2004}\natexlab{}.
\newblock \showarticletitle{Legalizing a {P}lacement with {M}inimum {T}otal
  {M}ovement}.
\newblock \bibinfo{journal}{\emph{TCAD}} \bibinfo{volume}{23},
  \bibinfo{number}{12} (\bibinfo{year}{2004}), \bibinfo{pages}{1597--1613}.
\newblock


\bibitem[\protect\citeauthoryear{Bustany, Chinnery, Shinnerl, and
  Yutsis}{Bustany et~al\mbox{.}}{2015}]%
        {bustany2015ISPD}
\bibfield{author}{\bibinfo{person}{I. Bustany}, \bibinfo{person}{D. Chinnery},
  \bibinfo{person}{J. Shinnerl}, {and} \bibinfo{person}{V. Yutsis}.}
  \bibinfo{year}{2015}\natexlab{}.
\newblock \showarticletitle{{I}{S}{P}{D} 2015 benchmarks with fence regions and
  routing blockages for detailed-routing-driven placement}. In
  \bibinfo{booktitle}{\emph{Proceedings of the ISPD}}.
  \bibinfo{pages}{157--164}.
\newblock


\bibitem[\protect\citeauthoryear{{Chen}, {Zhu}, {Zhu}, and {Chang}}{{Chen}
  et~al\mbox{.}}{2017}]%
        {QPLegalization2017}
\bibfield{author}{\bibinfo{person}{J. {Chen}}, \bibinfo{person}{Z. {Zhu}},
  \bibinfo{person}{W. {Zhu}}, {and} \bibinfo{person}{Y. {Chang}}.}
  \bibinfo{year}{2017}\natexlab{}.
\newblock \showarticletitle{Toward {O}ptimal {L}egalization for
  {M}ixed-{C}ell-{H}eight {C}ircuit {D}esigns}. In
  \bibinfo{booktitle}{\emph{54th DAC}}. 6.
\newblock
\showISBNx{9781450349277}


\bibitem[\protect\citeauthoryear{Cheng, Huang, Mak, and Wang}{Cheng
  et~al\mbox{.}}{2018}]%
        {SpacingRules}
\bibfield{author}{\bibinfo{person}{Y. Cheng}, \bibinfo{person}{D. Huang},
  \bibinfo{person}{W. Mak}, {and} \bibinfo{person}{T. Wang}.}
  \bibinfo{year}{2018}\natexlab{}.
\newblock \showarticletitle{A {P}ractical {D}etailed {P}lacement {A}lgorithm
  under {M}ulti-{C}ell {S}pacing {C}onstraints}. In
  \bibinfo{booktitle}{\emph{Proceedings of the ICCAD}}. 8.
\newblock


\bibitem[\protect\citeauthoryear{{Chow}, {Pui}, and {Young}}{{Chow}
  et~al\mbox{.}}{2016}]%
        {ChowPuiYoung}
\bibfield{author}{\bibinfo{person}{W. {Chow}}, \bibinfo{person}{C. {Pui}},
  {and} \bibinfo{person}{E. {Young}}.} \bibinfo{year}{2016}\natexlab{}.
\newblock \showarticletitle{Legalization {A}lgorithm for {M}ultiple-{R}ow
  {H}eight {S}tandard {C}ell {D}esign}. In \bibinfo{booktitle}{\emph{53rd
  DAC}}. \bibinfo{pages}{1--6}.
\newblock


\bibitem[\protect\citeauthoryear{{Darav}, {Bustany}, {Kennings}, and
  {Mamidi}}{{Darav} et~al\mbox{.}}{2017}]%
        {benchmarks}
\bibfield{author}{\bibinfo{person}{N. {Darav}}, \bibinfo{person}{I. {Bustany}},
  \bibinfo{person}{A. {Kennings}}, {and} \bibinfo{person}{R. {Mamidi}}.}
  \bibinfo{year}{2017}\natexlab{}.
\newblock \showarticletitle{{ICCAD}-2017 {CAD} {C}ontest in {M}ulti-{D}eck
  {S}tandard {C}ell {L}egalization and {B}enchmarks}. In
  \bibinfo{booktitle}{\emph{ICCAD}}. \bibinfo{pages}{867--871}.
\newblock


\bibitem[\protect\citeauthoryear{Garey and Johnson}{Garey and Johnson}{1978}]%
        {GareyJohnson}
\bibfield{author}{\bibinfo{person}{M. Garey} {and} \bibinfo{person}{D.
  Johnson}.} \bibinfo{year}{1978}\natexlab{}.
\newblock \showarticletitle{``{S}trong''\space{N}{P}-{C}ompleteness {R}esults:
  {M}otivation, {E}xamples, and {I}mplications}.
\newblock \bibinfo{journal}{\emph{J. ACM}} \bibinfo{volume}{25},
  \bibinfo{number}{3} (\bibinfo{year}{1978}), \bibinfo{pages}{499--508}.
\newblock


\bibitem[\protect\citeauthoryear{Garey, Tarjan, and Wilfong}{Garey
  et~al\mbox{.}}{1988}]%
        {SchedulingGareyTarjanWilfong}
\bibfield{author}{\bibinfo{person}{M. Garey}, \bibinfo{person}{R. Tarjan},
  {and} \bibinfo{person}{G. Wilfong}.} \bibinfo{year}{1988}\natexlab{}.
\newblock \showarticletitle{One-Processor Scheduling with Symmetric Earliness
  and Tardiness Penalties}.
\newblock \bibinfo{journal}{\emph{Mathematics of Operations Research}}
  \bibinfo{volume}{13}, \bibinfo{number}{2} (\bibinfo{year}{1988}),
  \bibinfo{pages}{330--348}.
\newblock


\bibitem[\protect\citeauthoryear{{Han}, {Han}, {Kahng}, {Lee}, {Wang}, and
  {Xu}}{{Han} et~al\mbox{.}}{2017}]%
        {DiffusionSteps}
\bibfield{author}{\bibinfo{person}{C. {Han}}, \bibinfo{person}{K. {Han}},
  \bibinfo{person}{A. {Kahng}}, \bibinfo{person}{H. {Lee}}, \bibinfo{person}{L.
  {Wang}}, {and} \bibinfo{person}{B. {Xu}}.} \bibinfo{year}{2017}\natexlab{}.
\newblock \showarticletitle{Optimal {M}ulti-{R}ow {D}etailed {P}lacement for
  {Y}ield and {M}odel-{H}ardware {C}orrelation {I}mprovements in {S}ub-10nm
  {V}{L}{S}{I}}. In \bibinfo{booktitle}{\emph{ICCAD}}.
  \bibinfo{pages}{667--674}.
\newblock


\bibitem[\protect\citeauthoryear{{Han}, {Kahng}, {Wang}, and {Xu}}{{Han}
  et~al\mbox{.}}{2019}]%
        {DiffusionStepsEnhanced}
\bibfield{author}{\bibinfo{person}{C. {Han}}, \bibinfo{person}{A. {Kahng}},
  \bibinfo{person}{L. {Wang}}, {and} \bibinfo{person}{B. {Xu}}.}
  \bibinfo{year}{2019}\natexlab{}.
\newblock \showarticletitle{Enhanced {O}ptimal {M}ulti-{R}ow {D}etailed
  {P}lacement for {N}eighbor {D}iffusion {E}ffect {M}itigation in {S}ub-10 nm
  {V}{L}{S}{I}}.
\newblock \bibinfo{journal}{\emph{TCAD}} \bibinfo{volume}{38},
  \bibinfo{number}{9} (\bibinfo{year}{2019}), \bibinfo{pages}{1703--1716}.
\newblock


\bibitem[\protect\citeauthoryear{Hill}{Hill}{2002}]%
        {hill}
\bibfield{author}{\bibinfo{person}{D. Hill}.} \bibinfo{year}{2002}\natexlab{}.
\newblock \bibinfo{title}{Method and system for high speed detailed placement
  of cells within an integrated circuit design}.
\newblock \bibinfo{howpublished}{U.S. Patent 6370673}.
\newblock


\bibitem[\protect\citeauthoryear{Hung, Chou, and Mak}{Hung
  et~al\mbox{.}}{2017}]%
        {hung2017mixed}
\bibfield{author}{\bibinfo{person}{C. Hung}, \bibinfo{person}{P. Chou}, {and}
  \bibinfo{person}{W. Mak}.} \bibinfo{year}{2017}\natexlab{}.
\newblock \showarticletitle{Mixed-{C}ell-{H}eight {S}tandard {C}ell {P}lacement
  {L}egalization}. In \bibinfo{booktitle}{\emph{Proceedings of the Great Lakes
  Symposium on VLSI}}. \bibinfo{pages}{149--154}.
\newblock


\bibitem[\protect\citeauthoryear{{Kahng}, {Tucker}, and {Zelikovsky}}{{Kahng}
  et~al\mbox{.}}{1999}]%
        {KahngTuckerZelikovsky}
\bibfield{author}{\bibinfo{person}{A. {Kahng}}, \bibinfo{person}{P. {Tucker}},
  {and} \bibinfo{person}{A. {Zelikovsky}}.} \bibinfo{year}{1999}\natexlab{}.
\newblock \showarticletitle{Optimization of {L}inear {P}lacements for
  {W}irelength {M}inimization with {F}ree {S}ites}. In
  \bibinfo{booktitle}{\emph{Proceedings of the Asia and South Pacific Design
  Automation Conference}}. \bibinfo{pages}{241--244}.
\newblock


\bibitem[\protect\citeauthoryear{Korte, Rautenbach, and Vygen}{Korte
  et~al\mbox{.}}{2007}]%
        {BonnTools}
\bibfield{author}{\bibinfo{person}{B. Korte}, \bibinfo{person}{D. Rautenbach},
  {and} \bibinfo{person}{J. Vygen}.} \bibinfo{year}{2007}\natexlab{}.
\newblock \showarticletitle{{B}onn{T}ools: {M}athematical {I}nnovation for
  {L}ayout and {T}iming {C}losure of {S}ystems on a {C}hip}.
\newblock \bibinfo{journal}{\emph{Proc. IEEE}}  \bibinfo{volume}{95}
  (\bibinfo{year}{2007}), \bibinfo{pages}{555--572}.
\newblock


\bibitem[\protect\citeauthoryear{{Li}, {Chow}, {Chen}, {Young}, and {Yu}}{{Li}
  et~al\mbox{.}}{2018}]%
        {Routability}
\bibfield{author}{\bibinfo{person}{H. {Li}}, \bibinfo{person}{W. {Chow}},
  \bibinfo{person}{G. {Chen}}, \bibinfo{person}{E. {Young}}, {and}
  \bibinfo{person}{B. {Yu}}.} \bibinfo{year}{2018}\natexlab{}.
\newblock \showarticletitle{Routability-Driven and Fence-Aware Legalization for
  Mixed-Cell-Height Circuits}. In \bibinfo{booktitle}{\emph{55th DAC}}.
  \bibinfo{pages}{1--6}.
\newblock


\bibitem[\protect\citeauthoryear{Li, Chen, Zhu, and Chang}{Li
  et~al\mbox{.}}{2019}]%
        {Analytical}
\bibfield{author}{\bibinfo{person}{X. Li}, \bibinfo{person}{J. Chen},
  \bibinfo{person}{W. Zhu}, {and} \bibinfo{person}{Y. Chang}.}
  \bibinfo{year}{2019}\natexlab{}.
\newblock \showarticletitle{Analytical Mixed-Cell-Height Legalization
  Considering Average and Maximum Movement Minimization}. In
  \bibinfo{booktitle}{\emph{Proceedings of the ISPD}}. \bibinfo{pages}{27--34}.
\newblock


\bibitem[\protect\citeauthoryear{{Lin}, {Yu}, {Xu}, {Gao}, {Viswanathan},
  {Liu}, {Li}, {Alpert}, and {Pan}}{{Lin} et~al\mbox{.}}{2016}]%
        {MrDP}
\bibfield{author}{\bibinfo{person}{Y. {Lin}}, \bibinfo{person}{B. {Yu}},
  \bibinfo{person}{X. {Xu}}, \bibinfo{person}{J. {Gao}}, \bibinfo{person}{N.
  {Viswanathan}}, \bibinfo{person}{W. {Liu}}, \bibinfo{person}{Z. {Li}},
  \bibinfo{person}{C. {Alpert}}, {and} \bibinfo{person}{D. {Pan}}.}
  \bibinfo{year}{2016}\natexlab{}.
\newblock \showarticletitle{Mr{D}{P}: {M}ultiple-row {D}etailed {P}lacement of
  {H}eterogeneous-sized {C}ells for {A}dvanced {N}odes}. In
  \bibinfo{booktitle}{\emph{ICCAD}}. \bibinfo{pages}{1--8}.
\newblock


\bibitem[\protect\citeauthoryear{Spindler, Schlichtmann, and Johannes}{Spindler
  et~al\mbox{.}}{2008}]%
        {Abacus}
\bibfield{author}{\bibinfo{person}{P. Spindler}, \bibinfo{person}{U.
  Schlichtmann}, {and} \bibinfo{person}{F. Johannes}.}
  \bibinfo{year}{2008}\natexlab{}.
\newblock \showarticletitle{Abacus: {F}ast {L}egalization of {S}tandard {C}ell
  {C}ircuits with {M}inimal {M}ovement}. In
  \bibinfo{booktitle}{\emph{Proceedings of the ISPD}}. \bibinfo{pages}{47--53}.
\newblock


\bibitem[\protect\citeauthoryear{Suhl}{Suhl}{2010}]%
        {Suhl}
\bibfield{author}{\bibinfo{person}{U. Suhl}.} \bibinfo{year}{2010}\natexlab{}.
\newblock \emph{\bibinfo{title}{Row-{Placement in {V}{L}{S}{I} Design: The
  Clumping Algorithm and a generalization}}}.
\newblock diploma thesis. \bibinfo{school}{University of Bonn},
  \bibinfo{address}{Research Institute for Discrete Mathematics}.
\newblock


\bibitem[\protect\citeauthoryear{Tarjan}{Tarjan}{1983}]%
        {Tarjan}
\bibfield{author}{\bibinfo{person}{R. Tarjan}.}
  \bibinfo{year}{1983}\natexlab{}.
\newblock \bibinfo{booktitle}{\emph{Data Structures and Network Algorithms}}.
\newblock \bibinfo{publisher}{SIAM}.
\newblock
\showISBNx{0898711878}


\bibitem[\protect\citeauthoryear{{Wang}, {Wu}, {Chen}, {Chang}, {Kuo}, {Zhu},
  and {Fan}}{{Wang} et~al\mbox{.}}{2017}]%
        {EffectiveLegalizationAlgorithm}
\bibfield{author}{\bibinfo{person}{C. {Wang}}, \bibinfo{person}{Y. {Wu}},
  \bibinfo{person}{J. {Chen}}, \bibinfo{person}{Y. {Chang}},
  \bibinfo{person}{S. {Kuo}}, \bibinfo{person}{W. {Zhu}}, {and}
  \bibinfo{person}{G. {Fan}}.} \bibinfo{year}{2017}\natexlab{}.
\newblock \showarticletitle{An {E}ffective {L}egalization {A}lgorithm for
  {M}ixed-{C}ell-{H}eight {S}tandard {C}ells}. In
  \bibinfo{booktitle}{\emph{22nd Asia and South Pacific Design Automation
  Conference}}. \bibinfo{pages}{450--455}.
\newblock


\bibitem[\protect\citeauthoryear{Wu and Chu}{Wu and Chu}{2015}]%
        {WuChu}
\bibfield{author}{\bibinfo{person}{G. Wu} {and} \bibinfo{person}{C. Chu}.}
  \bibinfo{year}{2015}\natexlab{}.
\newblock \showarticletitle{Detailed {P}lacement {A}lgorithm for {V}{L}{S}{I}
  {D}esign with {D}ouble-{R}ow {H}eight {S}tandard {C}ells}.
\newblock \bibinfo{journal}{\emph{TCAD}}  \bibinfo{volume}{35}
  (\bibinfo{year}{2015}), \bibinfo{pages}{1569--1573}.
\newblock


\bibitem[\protect\citeauthoryear{Zhu, Li, Chen, Chen, Zhu, and Chang}{Zhu
  et~al\mbox{.}}{2018}]%
        {QPLegalization2018}
\bibfield{author}{\bibinfo{person}{Z. Zhu}, \bibinfo{person}{X. Li},
  \bibinfo{person}{Y. Chen}, \bibinfo{person}{J. Chen}, \bibinfo{person}{W.
  Zhu}, {and} \bibinfo{person}{Y. Chang}.} \bibinfo{year}{2018}\natexlab{}.
\newblock \showarticletitle{{M}ixed-{C}ell-{H}eight {L}egalization
  {Co}nsidering {T}echnology and {R}egion {C}onstraints}. In
  \bibinfo{booktitle}{\emph{Proceedings of the ICCAD}}. 8.
\newblock


\end{thebibliography}

\appendix

\end{document}